\documentclass{llncs}
\usepackage[latin1]{inputenc}
\usepackage[english]{babel}
\usepackage{amsmath}
\usepackage{amsfonts}
\usepackage{amssymb}
\usepackage{graphicx}
\usepackage{subfigure}
\usepackage{tikz}

\usetikzlibrary{matrix}

\author{Alexander L\"uck\inst{1} \and Verena Wolf\inst{1}}
\institute{Department of Computer Science, Saarland University, Saarbr\"ucken, Germany
}

\title{A Stochastic Automata Network Description for Spatial DNA-Methylation Models}
\begin{document}
\maketitle

\begin{abstract}
 DNA methylation is an important biological mechanism to regulate gene expression and control cell development. 
 Mechanistic modeling has become a popular approach to enhance our understanding of the dynamics of methylation pattern formation in living cells. 
 Recent findings suggest that the methylation state of a cytosine base can be influenced by its DNA neighborhood. 
 Therefore, it is necessary to generalize existing mathematical models that consider only one cytosine and its partner on the opposite DNA-strand (CpG), in order to include such neighborhood dependencies. 
 One approach is to describe the system as a stochastic automata network (SAN) with functional transitions. We show that single-CpG models can successfully be generalized to multiple CpGs using the SAN description and verify the results by comparing them to results from extensive Monte-Carlo simulations.
\end{abstract}

\section{Introduction}
 The field of epigenetics investigates  changes in gene function that are not related to the organism's genetic code but rely on epigenetic markers.
 DNA methylation is one of the main epigenetic markers regulating gene expression and controlling cell fate. 
 The epigenetic code is passed along from one generation to the next but environmental factors such as diet, stress, and prenatal nutrition can modify  such markers resulting, for instance, in the deactivation or activation of certain genes.
 In the case of  DNA methylation, a marker is set by adding a methyl group to cytosine (C) bases on the DNA. 
The conversion of C to its methylated form (5-methylcytosine) is carried out by DNA methyltransferase (Dnmt) enzymes and the resulting methylation pattern 
determines the program and function of the corresponding cell.

Together with a subsequent guanine (G) and the corresponding GC-pair on the opposite strand, a cytosine base forms a so-called CpG, which may be methylated on none, one, or both DNA strands.
Mechanistic models based on Discrete-Time Markov Chains (DTMCs) have been developed to describe the temporal evolution of the methylation state of a CpG \cite{arand2012vivo}.
The goal of such models is to increase the understanding of methylation pattern formation. 
%While each individual CpG may undergo the same modifications of its methylation state, independent of its neighboring DNA bases, the rate at which such transitions occur may depend on the neighborhood. 
Due to the possible influence of the neighboring CpGs \cite{haerter2013collaboration,lovkvist2016dna}   generalizations of   simple, single-CpG  models have recently been developed, which take more than one CpG into account \cite{bonello2013bayesian,luck2019hidden,luck2017stochastic,meyer2017modeling}.
However, even for a small number of CpGs the size of the transition matrices grows rapidly and they are therefore hard to generate without suitable methods.
One of these methods are the stochastic automata networks (SANs) \cite{plateau1991stochastic,stewart1995numerical}.
In this work we show how to successfully apply the SAN framework to Markov models for single CpGs in order to generalize them to multiple CpGs with the aid of suitable, neighbor state dependent rate functions.

This paper is organized is follows:
In Section~2 we describe the model for a single CpG and show how to generalize it to several CpGs using the SAN framework. 
A comparison with Monte-Carlo simulations in order to verify the results of our generated matrices can be found in Section~3.
In Section~4 we conclude our findings and give ideas for possible future work.

\section{SANS}
In order to describe the methylation dynamics of a CpG in terms of DTMCs, we consider  CpGs on the double stranded DNA where each CpG contains two Cs (one on each strand), each of which can be either methylated or unmethylated.
Therefore there are four possible states, i.e., both Cs unmethylated (state 1), only the C on the upper strand methylated (state 2), only the C on the lower strand methylated (state 3) and both Cs methylated (state 4).
The processes that lead to state transitions are cell division, maintenance and \emph{de novo} methylation. 

During cell division, one strand and its methylation state are kept as they are (parental strand), while the other strand is newly synthesized (daughter strand) with all Cs unmethylated.
Maintenance methylation adds methylation to the C on the daughter strand  if the corresponding C on the parental strand is already methylated in order to reestablish existing methylation patterns.
In contrast to that, \emph{de novo} methylation may occur on unmethylated Cs on both strands  independent of the methylation state of the C on the opposite strand and is therefore responsible for creating new patterns.
The transition probability matrices for these processes for a single CpG are listed in Tab.~\ref{Tab:TransMatr}.
In the left column the matrices concerning the upper strand are listed and in the right column the matrices for the lower strand.

It is biologically plausible to assume that cell division happens first, afterwards maintenance on the daughter strand and in the end \emph{de novo} on both strands takes place.
Since the strand that is kept after cell division is chosen randomly with equal probability, the total transition probability matrix $P$ is given by
\begin{equation}
P=0.5\cdot(CD_1\cdot M_1+CD_2\cdot M_2)\cdot T_1\cdot T_2.
\label{Eq:P1}
\end{equation}

\begin{figure}[tb]
\begin{center}
\vspace*{-1.5ex}
\includegraphics[scale=0.4]{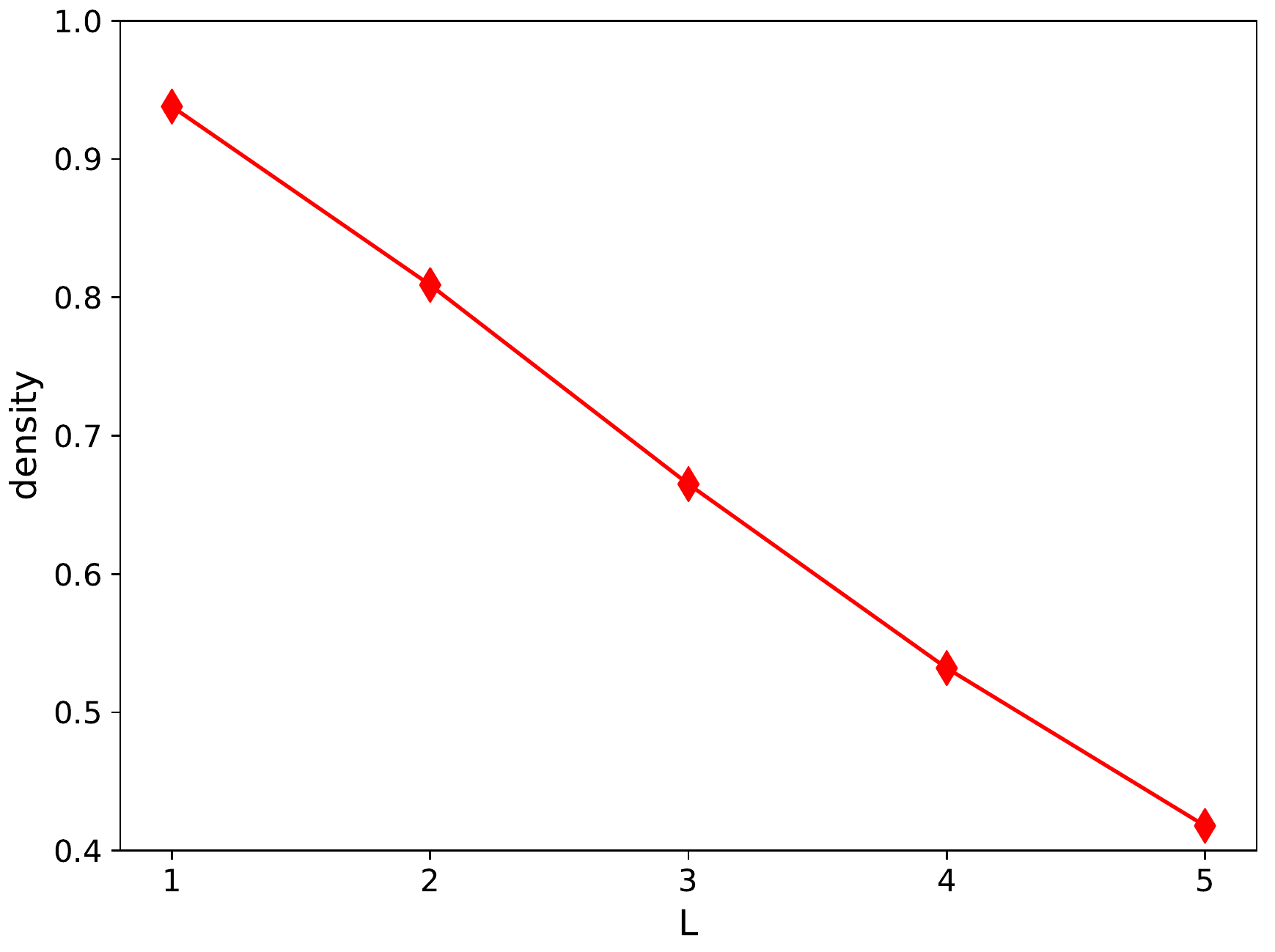}\\[-3ex]
\caption{Density of entries for the transition matrix $P$ for different numbers of CpGs $L$.\label{Fig:density}}
\end{center}
\vspace*{-3ex}
\end{figure}

\renewcommand{\arraystretch}{1.125}
\begin{table}[tb]
\begin{center}
\caption{Transition matrices for a single CpG. Note that the transition probabilities $f$ may be functions of the reaction parameters, the CpG position and/or the states of the adjacent CpGs. The matrices in the left column represent the transitions on the upper and the matrices in the right column the transitions on the lower strand. \label{Tab:TransMatr}}
\begin{tabular}{c c}
\multicolumn{2}{c}{\textbf{Cell Division}} \\[0.5ex]
$CD_1=\begin{pmatrix}
1 & 0 & 0 & 0 \\
1 & 0 & 0 & 0 \\
0 & 0 & 1 & 0 \\
~~0~~ & ~~0~~ & ~~1~~ & ~~0~~
\end{pmatrix}$ ~~~&~~~
$CD_2=\begin{pmatrix}
1 & 0 & 0 & 0 \\
0 & 1 & 0 & 0 \\
1 & 0 & 0 & 0 \\
~~0~~ & ~~1~~ & ~~0~~ & ~~0~~
\end{pmatrix}$ \\[2ex]
\multicolumn{2}{c}{\textbf{Maintenance}}\\[0.5ex]
$M_1=\begin{pmatrix}
1 & 0 & 0 & 0 \\
0 & 1 & 0 & 0 \\
0 & 0 & 1\!-\!f & f \\
~~0~~ & ~~0~~ & ~~0~~ & ~~1~~
\end{pmatrix}$ ~~~&~~~
$M_2=\begin{pmatrix}
1 & 0 & 0 & 0 \\
0 & 1\!-\!f & 0 & f \\
0 & 0 & 1 & 0 \\
~~0~~ & ~~0~~ & ~~0~~ & ~~1~~
\end{pmatrix}$ \\[2ex]
\multicolumn{2}{c}{\textbf{\emph{De Novo}}}\\[0.5ex]
$T_1=\begin{pmatrix}
1\!-\!f & f & 0 & 0 \\
0 & 1 & 0 & 0 \\
0 & 0 & 1\!-\!f & f \\
~~0~~ & ~~0~~ & ~~0~~ & ~~1~~
\end{pmatrix}$ ~~~&~~~
$T_2=\begin{pmatrix}
1\!-\!f & 0 & f & 0 \\
0 & 1\!-\!f & 0 & f \\
0 & 0 & 1 & 0 \\
~~0~~ & ~~0~~ & ~~0~~ & ~~1~~
\end{pmatrix}$\\[-6ex]
\end{tabular}
\end{center}
\end{table}
Given a sequence of $L$ CpGs, each CpG can be described by the aforementioned DTMC, which gives us one automaton of the automata network.
The structure of each automaton is independent of the automata describing neighboring CpGs, however, the transition probabilities may depend on 
their local states (functional transitions).
A suitable method to combine these automata in order to capture the dynamics of whole sequences of CpGs is to consider them as an automata network.
Within the SAN framework, the transition matrices of the individual automata are combined via the Kronecker product.
Since in our case the transition matrix for one automaton is a product of the transition matrices for the different processes, we exploit some properties of the Kronecker product to generate the global transition matrix of the network.
From\\[-4ex]
\begin{align}
A\otimes(B\otimes C)&=(A\otimes B)\otimes C  \label{Eq:assoc}\\
(AC) \otimes (BD)&= (A \otimes B) \cdot (C\otimes D) \label{Eq:MatrKron}
\end{align} \\[-4ex]
the following properties can be derived \cite{davio1981kronecker} \\[-4ex]
\begin{align}
\left(\prod_{n=1}^N A_n\right)\otimes \left(\prod_{n=1}^N B_n\right)&=\prod_{n=1}^N \left(A_n\otimes B_n\right), \label{Eq:kprod1}\\
\bigotimes^{M}_{m=1}\left(\prod_{n=1}^N A^{(m)}_n\right)&=\prod_{n=1}^N\left(\bigotimes^{M}_{m=1} A_n^{(m)} \right) \label{Eq:kprodM}.
\end{align} \\[-3ex]
%The proofs of Eqs.~\eqref{Eq:kprod1} and \eqref{Eq:kprodM} can be found in the Appendix.
Note that in Eqs.~\eqref{Eq:MatrKron}-\eqref{Eq:kprodM} the corresponding matrices have to be compatible under the standard matrix product.\\
As a consequence of Eq.~\eqref{Eq:kprodM} it is possible to obtain the total transition matrix $P$ in two ways: First compute a transition matrix for a single CpG (Eq.~\eqref{Eq:P1}) and extend the result to several CpGs with the Kronecker product or
  calculate the transition matrices for the different processes for several CpGs first via the Kronecker product and combine them afterwards.
Since the transition probabilities may depend on the neighbor states, i.e. the states of the adjacent automata, it is easier to choose the second possibility and construct the individual transition matrices for the different processes first.
Another advantage is that the matrices for the different processes are sparse, while the total transition matrix is quite dense for a single CpG (see Tab.~\ref{Tab:TransMatr} and Fig.~\ref{Fig:density}), such that we apply the Kronecker product to sparse matrices and multiply the (also sparse) results afterwards.\\
If we assume that all CpGs are methylated independent of their neighborhood, then no functional transitions are needed and the transition probabilities are constants.
The construction of the global transition matrix is then straightforward by simply applying the Kronecker product.
To model dependence, first observe that since only the transition probabilities, but not the transitions themselves, depend on the neighboring states, the structure of the global transition matrix is the same as in the independent case.
By using functions instead of constant probabilities, we are able to capture the effect of the neighbors on the transition rates. 
Another advantage of the functions is that we can incorporate different model assumptions (like processivity) by using different functions without altering the structure of the transition matrices.\\
To shape the function $f:=f(\vec{r},l, s_{l-1}, s_{l+1}) \in [0,1]$ in the matrices in Tab.~\ref{Tab:TransMatr} to our needs, we use the following inputs: 
\begin{itemize}
\item $\vec{r}$ is a vector with the reaction parameters,
\item $l \in\{1,\ldots,L\}$ is the position of the CpG such that boundary ($l=1,L$) and non-boundary ($l=2,\ldots, L-1$) CpGs can be distinguished,
\item $s_{l-1}$ is the state of the left neighboring CpG and 
\item $s_{l+1}$ is the state of the right neighboring CpG. 
\end{itemize} 
Depending on the methylation event (maintenance or \emph{de novo}) different para\-meter vectors $\vec{r}$ can be chosen.
Since in general all CpGs may undergo a reaction, the states of the neighboring CpGs that are used (before or after reaction) as an input for the function depends on the underlying assumptions.
This will be demonstrated in the following.

We first note that the indices of the matrices in Tab.~\ref{Tab:TransMatr} correspond to the states before and after transition, i.e. the entry $a_{i,j}$ corresponds to the probability of going from state $i$ to state $j$.
Furthermore, there is a unique relation between the indices of the initial matrices and the indices of the result of their Kronecker product \\[-4.5ex]
\begin{align}
a_{r,s}\cdot b_{v,w}&=(A\otimes B)_{p(r-1)+v, q(s-1)+w}, \label{Eq:KronInd}\\
(A\otimes B)_{i,j}&=a_{\lfloor(i-1)/p\rfloor+1,\lfloor(j-1)/q\rfloor+1}\cdot b_{i-\lfloor(i-1)/p\rfloor p,j-\lfloor(j-1)/q\rfloor q},
\end{align} \\[-4ex]
where $A$ is a $p\times q$- and $B$ an arbitrary matrix. 
These formulas can easily be generalized such that for a Kronecker product of $L$ matrices we know exactly the indices for each of the matrices and therefore the states before and after transition for each CpG.
We then use this knowledge to choose the correct transition probability depending on our assumptions.

For the transition probabilities, we impose, depending on the neighbor states, the following forms \cite{luck2017stochastic}: 
For \emph{de novo} methylation events (the state of the other strand does not matter) we have transition probabilities for non-boundary cases \\[-4.5ex]
\begin{align}
\circ\circ\circ&\rightarrow \circ\bullet\circ\qquad p_1=0.5\!\cdot\!(\psi_L+\psi_R)\tau, \label{pstart}\\
\bullet\circ\circ&\rightarrow \bullet\bullet\circ\qquad p_2=0.5\!\cdot\!(\psi_L+\psi_R)\tau+0.5\!\cdot\!(1-\psi_L),\\
\circ\circ\bullet&\rightarrow\circ\bullet\bullet\qquad p_3=0.5\!\cdot\!(\psi_L+\psi_R)\tau+0.5\!\cdot\!(1-\psi_R),\\
\bullet\circ\bullet&\rightarrow\bullet\bullet\bullet\qquad p_4=1-0.5\!\cdot\!(\psi_L+\psi_R)(1-\tau),
\end{align} \\[-4ex]
where an empty circle represents an unmethylated and a filled circle a methylated C.
The parameters $\psi_L$ and $\psi_R$ characterize the dependency on the neighbor state, where $\psi_i=1$ means fully independent and $\psi_i=0$  full dependency.
 $\tau$ corresponds to the \emph{de novo} probability in the independent case.

For maintenance events, we replace the probability $\tau$ by $\mu$ and have the additional requirement that the C on the opposite strand must be methylated.\\
For boundary cases,  i.e., the left- and rightmost CpG in a sequence of $L$ CpGs, we have the probabilities \\[-4.5ex]
\begin{align}
?\circ\circ&\rightarrow\text{?}\bullet\circ\qquad \tilde{p}_1=(1-\rho)\!\cdot\!p_1+\rho\!\cdot\!p_2,\\
?\circ\bullet&\rightarrow\text{?}\bullet\bullet\qquad \tilde{p}_2=(1-\rho)\!\cdot\!p_3+\rho\!\cdot\!p_4,\\
\circ\circ\text{?}&\rightarrow \circ\bullet\text{?}\qquad \tilde{p}_3=(1-\rho)\!\cdot\!p_1+\rho\!\cdot\!p_3,\\
\bullet\circ\text{?}&\rightarrow\bullet\bullet\text{?}\qquad \tilde{p}_4=(1-\rho)\!\cdot\!p_2+\rho\!\cdot\!p_4, \label{pend}
\end{align} \\[-4ex]
where we use the average methylation level $\rho$, since we do not have any information about the CpG on the left/right at the boundaries.

For a moderate number of CpGs ($\approx 5$) it is possible to explicitely construct the whole transition matrix with a simple algorithm.
We first note that we have to apply the Kronecker product for the matrices in Tab.~\ref{Tab:TransMatr} $L$ times with themselves for a sequence of $L$ CpGs. 
We then apply the following scheme:\\[-3.5ex]
\begin{enumerate}
\item Identify the indices of the non-zero entries of the matrix. 
\item Calculate the indices of the resulting matrix after applying the Kronecker product with Eq.~\eqref{Eq:KronInd} for the indices from step 1. Iteratively applying Eq.~\eqref{Eq:KronInd} $L$ times leads to the final indices $(u,v)$.
For each $(u,v)$ we get an ordered list $\ell$ containing the indices from the original matrices that lead to this index.
\item For each $(u,v)$ calculate the matrix entry\\[-2.5ex]
\begin{equation}
m_{u,v}=\prod_{(i,j)\in \ell} a_{i,j},
\end{equation}\\[-2.25ex]
where $a_{i,j}$ are the entries of the original matrix.
\item If $a_{i,j}$ contains the function $f$ choose the neighbor states based on the assumption and the indices (states) from $\ell$ of the adjacent matrices.\\[-4ex]
\end{enumerate}
Note that for real data we have to ensure that all CpGs of a given sequence originate from the same cell in order to properly investigate the neighborhood dependencies.
Real data rarely covers states of more than a couple of successive CpGs from the same cell with sufficiently deep coverage.
Therefore, the number of contiguous CpGs is usually very limited, such that the explicit construction of the transition matrix
for short CpG sequences is feasible in most cases. 
For a possible larger number of CpGs from advanced measurement techniques we have to resort to more sophisticated methods to obtain the transition matrices~\cite{buchholz1995equivalence,buchholz1995hierarchical,buchholz2004kronecker,stewart1995numerical}.
\\[1ex]
\textbf{Example: Processivity}~\\
Since detailed mechanisms about the interaction of the Dnmts with the DNA remain elusive, we would like to test different assumptions such as processive or distributive methylation  mechanisms
\cite{gowher2002molecular,holz2010inherent,luck2017stochastic,norvil2016dnmt3b}.   
For the remainder of this paper, we assume \emph{processivity from left to right}, which is a reasonable assumption for Dnmt1, due to its link to the replication machinery.
The processivity from left to right implies, that a transition already happened at the left neighbor (position $l-1$) but not yet at the right neighbor ($l+1$).
This means, given the list of indices $\ell=[\ldots, (i_{l-1},j_{l-1}),(i_l,j_l),(i_{l+1},j_{l+1}),\ldots ]$ we choose $j_{l-1}$ for the left neighbor state and $i_{l+1}$ for the right neighbor state as an input for the function in step 4 of our algorithmic scheme. \\[-5ex]

\section{Results}

\begin{figure}[tb]
\vspace*{-1.5ex}
\begin{subfigure}[(0.8,0,0,0.1)]{\includegraphics[width=0.45\textwidth]{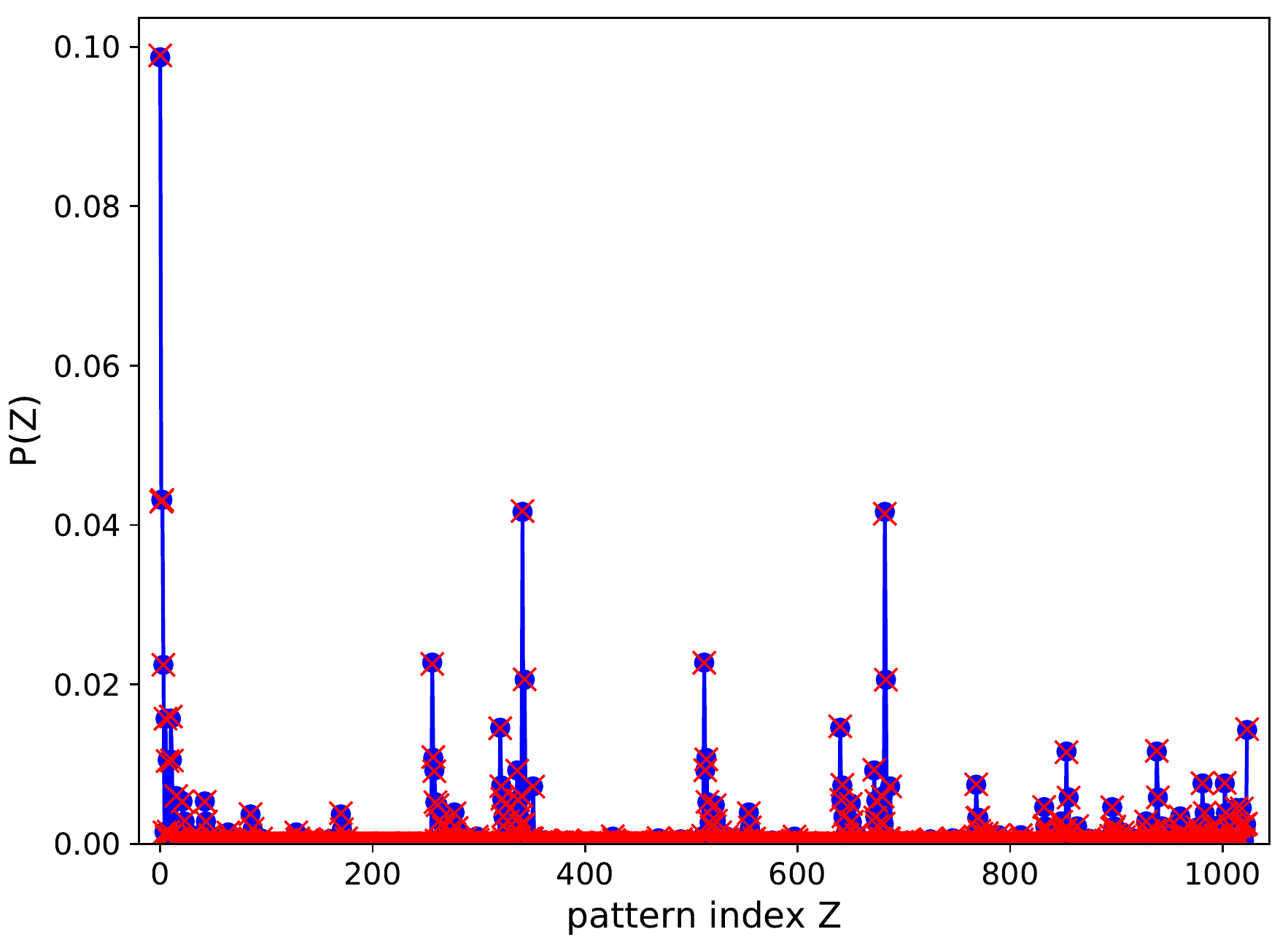}}
\end{subfigure}\hfill
\begin{subfigure}[Detailed excerpt from (a).]{\includegraphics[width=0.45\textwidth]{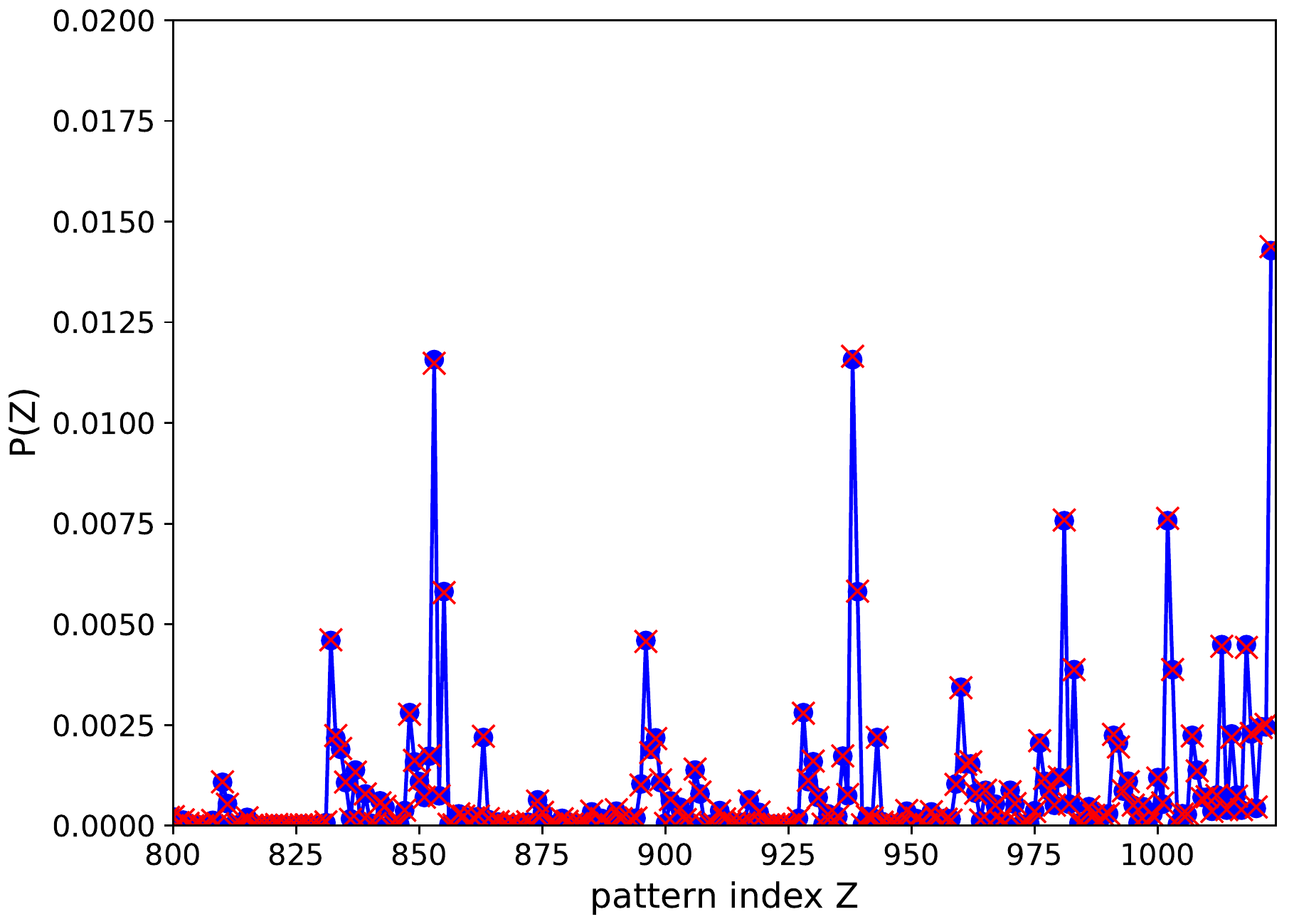}}
\end{subfigure}\\[-3ex]
\begin{subfigure}[(0.8,1,1,0.1)]{\includegraphics[width=0.45\textwidth]{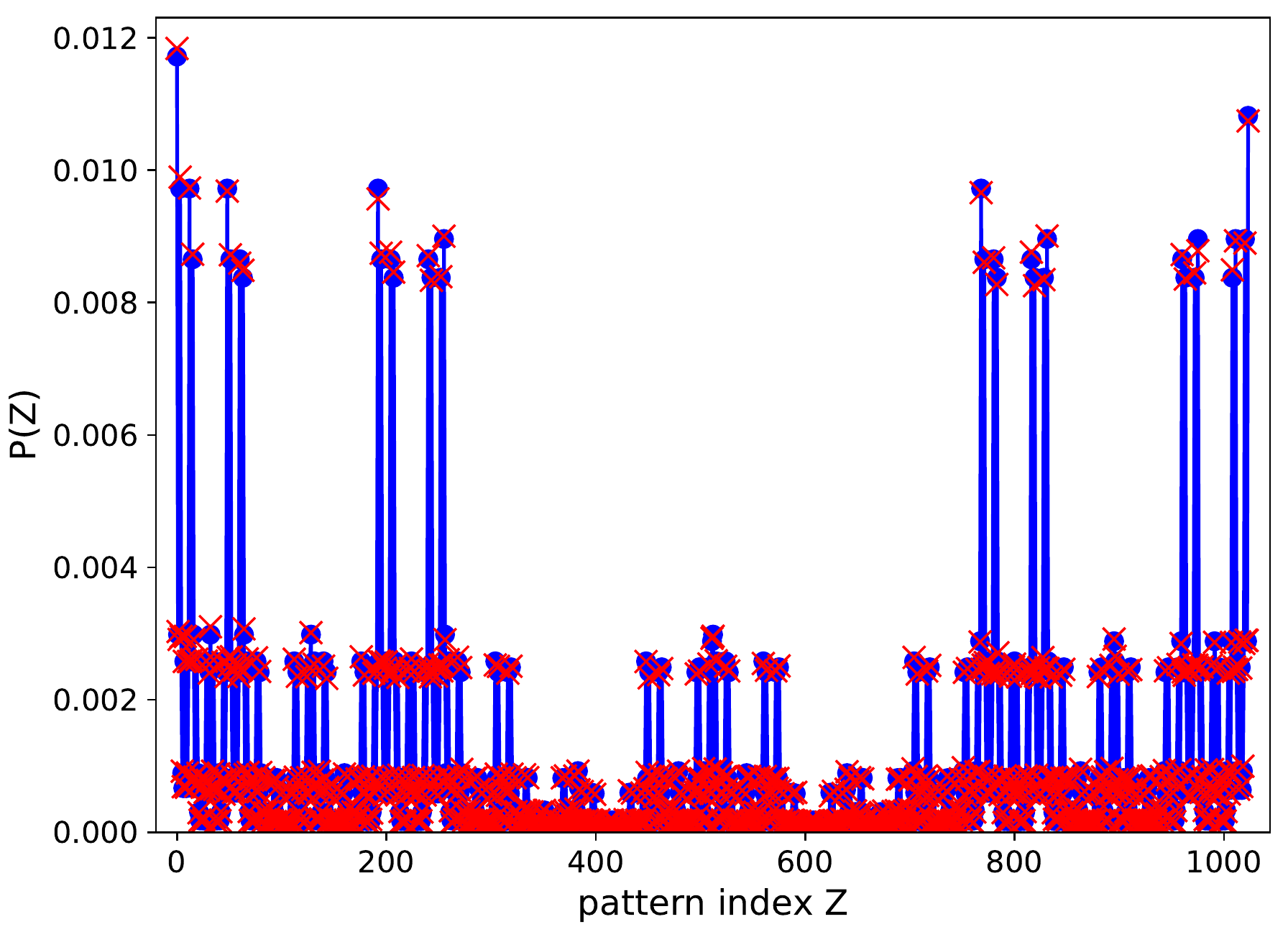}}
\end{subfigure}\hfill
\begin{subfigure}[(0.8,0.4,0.6,0.1)]{\includegraphics[width=0.45\textwidth]{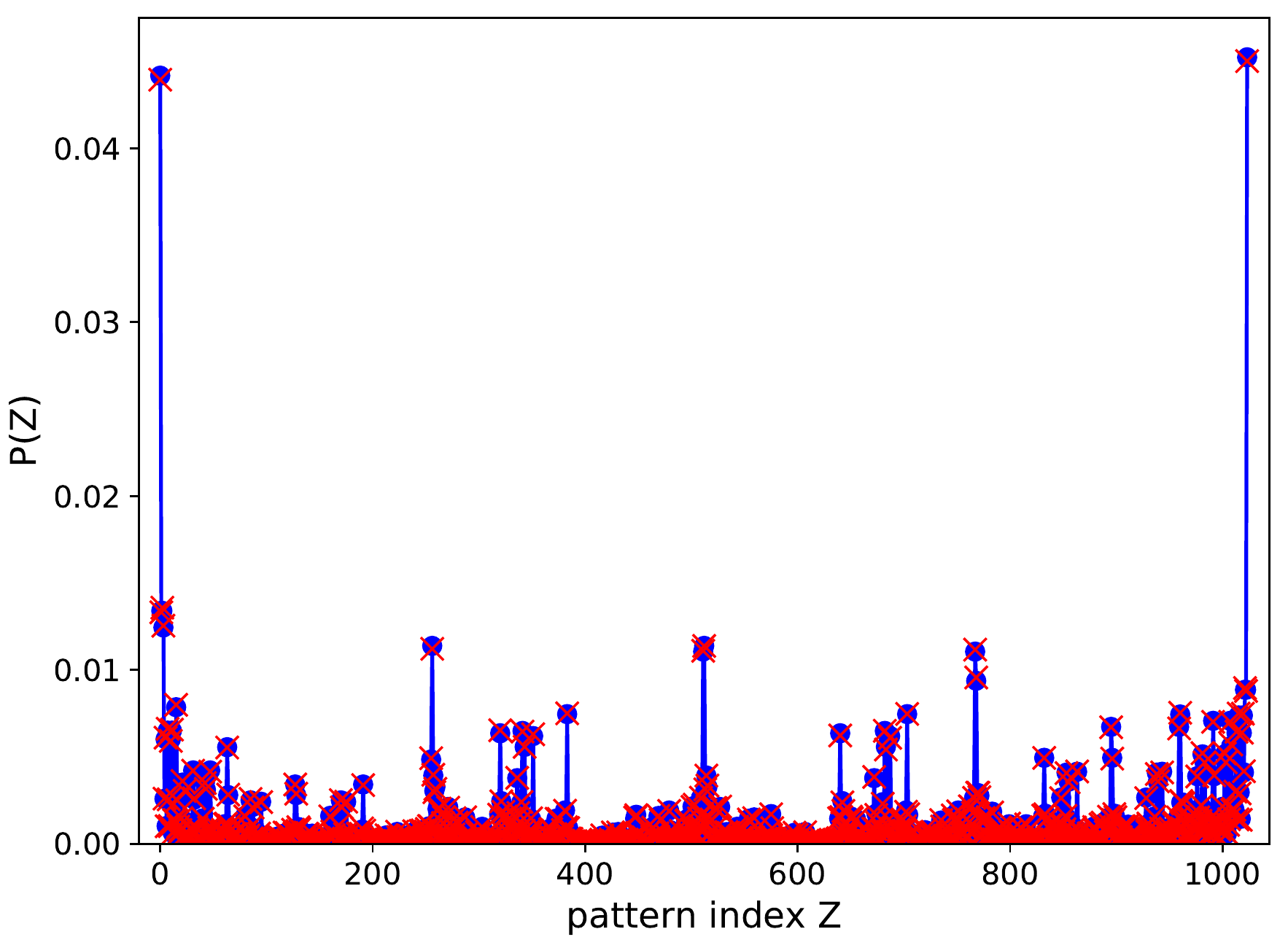}}
\end{subfigure}\\[-3ex]
\caption{Comparison of distributions obtained from transition matrices generated from the SAN description (blue) and from MC simulations (red). The parameters for each subfigure are given in the form $(\mu,\psi_L,\psi_R,\tau)$.\label{Fig:compMC}}
\vspace*{-3ex}
\end{figure}

In order to check the correctness of the matrices that we generate with the Kronecker product for $L$ CpGs, we compare the resulting distributions with results from Monte-Carlo (MC) simulations.
As initial distribution $\pi_0$ we use a discrete uniform distribution which assigns the same probability to all possible methylation patterns.
We then compute distributions $\pi(t)$ after $t=30$ cell division and subsequent methylation events via \\[-2ex]
\begin{equation}
\pi(t+1)=\pi(t)\cdot P,
\end{equation}\\[-3ex]
where $P$ is the total transition matrix, where we assume processivity.
We perform the corresponding MC simulations with $N=10^6$ runs.
The distributions for different parameter sets are shown in Fig.~\ref{Fig:compMC}.
Panels (a) and (b) show the fully dependent case, where the transition probabilities depend only on the neighbor states and not on the actual maintenance and \emph{de novo} rates $\mu$ and $\tau$.
In Fig.~\ref{Fig:compMC} (c) the transition probabilities are totally independent of the neighboring states and depend solely on $\mu$ and $\tau$.
This case is equivalent to the case where we replace the function $f$ by the respective (constant) transition probabilities.
Fig.~\ref{Fig:compMC} (d) shows a case with some dependency on both neighbor states, where the dependence to the left is slightly stronger.
Choosing a wrong transition function in the matrix entries (compared to MC, where it is easier to ensure the correct choice) would affect the distribution in (a) and (b) the most, since there is a full dependency and hence the largest effect from the neighboring states.
For the partial dependencies in (d) there should also be an effect if the choices were wrong.
In the independent case (c) there can not be a wrong choice, since the transition function is a constant.

\begin{figure}[tb]
\begin{center}
\vspace*{-1.5ex}
\includegraphics[scale=0.4]{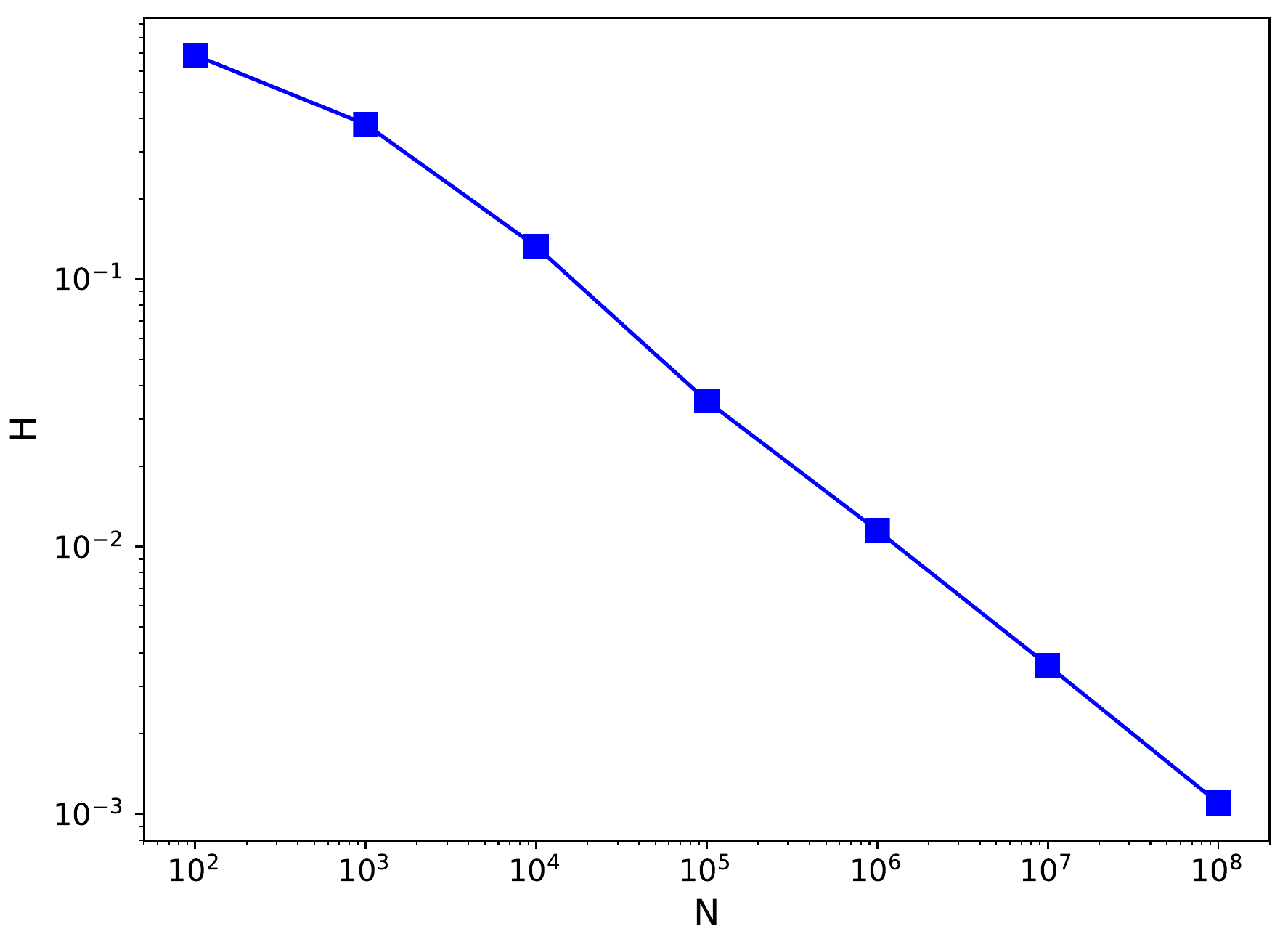}\\[-3ex]
\caption{Hellinger distance $H$ between distribution obtained from numerical SAN solution and from MC simulations with $N$ runs for the parameter set of Fig.~\ref{Fig:compMC}~(d).\label{Fig:hellinger}}
\end{center}
\vspace*{-7.5ex}
\end{figure}

In all cases we observe an almost perfect agreement with only small deviations on some patterns on a very small scale.
In order to exclude a flaw in the construction of the transition matrix we compute the Hellinger distance \\[-2ex]
\begin{equation}
 H(P,Q)=\frac {1}{\sqrt {2}} \left(\sum_{i=1}^{k}\left(\sqrt {p_{i}}-\sqrt {q_{i}}\right)^{2}\right)^{\frac{1}{2}}
\end{equation} \\[-2ex]
to compare the similarity of the distributions and to check if the deviations stem from the finite number of MC simulation runs.
From Fig.~\ref{Fig:hellinger} it is obvious that with an increasing number of runs the distributions become more and more similar such that we can indeed exclude a flaw in the matrix construction.
The small deviations stem from the finite number of runs since for $N=10^6$  there are still statistical inaccuracies and hence $H$ is quite large (order of $10^{-2}$).\\[-5ex]

\section{Conclusion}
In this paper, we propose an adaptation of the stochastic automata network framework to solve problems related to the generation of  transition probability  matrices for a biological application. With the proposed procedure we are able to describe the methylation dynamics of a sequence of several CpGs, given the matrices for a single CpG.
The transition matrices form the basis for numerical solutions of mechanistic models, which allow to test assumptions about the functioning of methylation enzymes.
The presented framework allows to consider longer CpG sequences and hence the proper investigation of larger genomic regions.
The transition matrix for $L$ CpGs can systematically be generated from the small single-CpG matrices and the generation is less prone to errors than an ad-hoc approach.
It is also pretty easy to adapt the model to test different biological assumptions by using different functions in the transition matrices.
In our example, we assumed processivity from left to right, but by changing the functions other assumptions like processivity from right to left (less biologically plausible) or even non-processive (e.g. distributive) behavior can be realized.
It is also easily possible to introduce additional reaction parameters for each individual CpG within this framework to generalize the model. Using the same reaction parameters for all CpGs is a strong assumption, especially for the neighborhood dependencies, which should intuitively be different due to the (in general) different distances between CpGs or also due to different base sequences in the DNA.\\
%It is also possible to generalize the functions and take more the states of more CpGs in the neighborhood into account.\\
Furthermore, it is also straightforward to apply the SAN approach to more complex methylation models in order to investigate possible neighborhood dependencies.
This is especially useful when there are more than four states per CpG as with more states the transition matrix grows rapidly.
Introducing additional hydroxylated Cs as in \cite{giehr2016influence} the number of states per CpG grows from four to eight such that the transition matrix grows from $4^L\times4^L$ to $8^L\times 8^L$ for $L$ CpGs.\\
With even more possible modifications of C, such as the formylated form 5-formylcyto\-sin, the number of possible states and hence the matrix size grows even more ($16^L$ or $16^L\times 16^L$ respectively).
In this case, the SAN description becomes even more useful as it would be very tedious to generate the transition matrix in other ways. 
It is also possible to apply the SAN approach to continuous time Markov chains or hybrid models as in \cite{kyriakopoulos2019hybrid}.
Here, the discrete transition matrix was generated with a Kronecker product, while 
the continuous generator matrix can be generated with a Kronecker sum.

\bibliographystyle{splncs03}

\end{document}